\documentclass[a4paper,10pt]{article}
\usepackage{tikz, amsmath}
\usetikzlibrary{calc}
\usepackage{bbm}
\usepackage{cite}
\usepackage{feynman}
\usepackage{xcolor}
\usepackage{dsfont}
\usepackage{subfigure}
\usepackage{authblk}
\usepackage{hyperref}
\usepackage{microtype}
\usepackage{caption}
\usepackage{amsmath}
\usepackage{float}
\usepackage{graphics}
\usepackage{psfrag}
\usepackage{color}
\usepackage{slashed}
\usepackage{subfigure}
\usepackage{graphicx}
\usepackage{array,multirow,axodraw}
\usepackage{ulem}
\usepackage{amsfonts}
\usepackage{amssymb}

\title{Clockwork for Neutrino Masses and Lepton Flavor Violation}
\author[a,b]{Alejandro Ibarra\thanks{ibarra@tum.de}}
\author[c]{Ashwani Kushwaha\thanks{ashwani@chep.iisc.ernet.in}}
\author[c]{Sudhir K.  Vempati\thanks{vempati@chep.iisc.ernet.in}}
\affil[a]{Physik-Department T30d, Technische Universit\"at M\"unchen,\protect \\James-Franck-Stra\ss{}e, 85748 Garching, Germany}
\affil[b]{School of Physics, Korea Institute for Advanced Study, Seoul 02455, South Korea}
\affil[c]{Centre for High Energy Physics, Indian Institute of Science  \protect \\
C. V. Raman Avenue, Bangalore 560012, India}
\usepackage[left=1in,top=1in,right=1in,bottom=1in,nohead]{geometry} 
\newcommand{\myname}[1]{%
	\begin{tikzpicture}[overlay, remember picture]
	\node[anchor=east] at ( $ (current page.north east) + (-1in,-2cm) $) {#1};
	\end{tikzpicture}%
}

\begin{document}
	\maketitle
	\myname{TUM-HEP 1107/17, KIAS-P17115}
\begin{abstract}
We investigate the generation of small neutrino masses in a clockwork framework which includes Dirac mass terms as well 
as Majorana mass terms for the new fermions. We derive analytic formulas for the masses of the new particles and for their 
Yukawa couplings to the lepton doublets, in the scenario where the clockwork parameters are universal. When the Majorana masses 
all vanish, the zero mode of the clockwork sector forms a Dirac pair with the active neutrino, with a mass which is in agreement 
with oscillations experiments for a sufficiently large number of clockwork gears. On the other hand, when the Majorana masses do
not vanish, neutrino masses are generated via the seesaw mechanism. In this case, and due to the fact that the effective Yukawa 
couplings of the higher modes can be sizable, neutrino masses can only be suppressed by postulating a large Majorana mass for 
all the gears. Finally, we discuss the constraints on the mass scale of the clockwork fermions from the non-observation of the 
rare leptonic decay $\mu\rightarrow e\gamma$. 
  
\end{abstract}
\newpage
\section{Introduction}
The smallness of neutrino masses stands as one of the most puzzling open questions in Fundamental Physics. 
A plausible solution to this puzzle is provided by the seesaw mechanism, in which the smallness of neutrino masses 
is explained by the breaking of the lepton number at a very high energy scale 
\cite{Minkowski:1977sc,Mohapatra:1979ia,Yanagida:1979as,GellMann:1980vs}. Models with conserved lepton number, 
on the other hand, can also reproduce the observations, at the expense of postulating tiny Yukawa couplings of 
the neutrino to the Standard Model Higgs. Such small parameters are usually regarded as unnatural, however the 
existence of tiny Yukawa couplings is a phenomenologically viable possibility, and can be accomplished in further 
extensions of the model (for reviews and recent models, 
see {\it e.g.} in~\cite{Mohapatra:2005wg, Wang:2017mcy, Krauss:2002px, Wang:2016lve, Ma:2016mwh, Borah:2017leo, Borah:2016zbd, Bonilla:2016diq, Kanemura:2016ixx, Okada:2014vla, Farzan:2012sa, Gu:2007ug, PhysRevLett.58.1600, PhysRevD.65.053014}).

Recently, a new mechanism of generating small couplings in theories
coupled to the Standard Model has been introduced~\cite{Choi:2015fiu, Kaplan:2015fuy}.  The mechanism, reminiscent of deconstruction 
models \cite{ArkaniHamed:2001ca, Hill:2000mu}, can 
be summarized as a linear quiver model with no large hierarchies 
in the theory parameters, that gives rise to site-dependent suppressed
couplings to the zero-mode \cite{Giudice:2016yja}. Originally, 
introduced for a quiver of Abelian Goldstone bosons (axions), it
has been generalized to fermions, vectors and other 
fields \cite{Giudice:2016yja,Craig:2017cda} (See also \cite{ Giudice:2017suc}). 
Applications and generalizations of this mechanism can be found 
in \cite{Saraswat:2016eaz, Kehagias:2016kzt, Farina:2016tgd, Ahmed:2016viu, Hambye:2016qkf, You:2017kah, Diez-Tejedor:2017ivd, Batell:2017kho, Tangarife:2017vnd, Coy:2017yex, Ben-Dayan:2017rvr, Hong:2017tel, Park:2017yrn, Lee:2017fin, Carena:2017qhd, Antoniadis:2017wyh,Ibanez:2017vfl}. 

In this work we explore the application of the fermionic clockwork to the generation of small neutrino masses. 
Concretely, we identify the right-handed neutrinos with the zero modes of a clockwork sector~\cite{Giudice:2016yja}, 
such that small couplings can be naturally generated and therefore small neutrino masses. We generalize the clockwork framework for the right
handed neutrinos by including also Majorana mass terms. We show that 
the clockwork mechanism, {\it i.e.}, the suppression of the Yukawa couplings
by site dependent power factors, is not affected by the presence 
of the Majorana mass terms. In fact, the combination of the clockwork
``suppression" and the Majorana ``seesaw" sets now the neutrino mass
scale.  When all the Majorana terms are set to zero, the 
clockwork provides an interesting alternative to the existing
models of Dirac neutrinos, which we investigate in this paper.
Furthermore, while the clockwork mechanism suppresses the couplings of the
zero mode, the couplings of the higher modes can be sizable and induce, via loops, potentially large rates for the leptonic rare decays.  

The rest of the paper is organized as follows. In section~\ref{neutrino}, 
we present the most general framework for clockwork neutrinos
with Dirac and Majorana mass terms, and we discuss their phenomenology in subsections ~\ref{Dirac} and~\ref{Majorana}, respectively. 
In section~\ref{LeptFlav}, we discuss lepton flavour violation 
in the clockwork scenario and calculate limits on the 
gear masses. We close with a summary. 

\section{Neutrinos in Clockwork }\label{neutrino}

We extend the Standard Model with $n$ left-handed and $n+1$ right-handed chiral fermions, 
singlets under the Standard Model gauge group, which we denote as $\psi_{Li}(i=0,...,n-1)$ and $\psi_{Ri}(i=0,...,n)$ respectively. 
The Lagrangian of the model reads:
\begin{align}
{\cal L}={\cal L}_{\rm SM}+{\cal L}_{\rm Clockwork}+{\cal L}_{\rm int}\;,
\end{align}
where ${\cal L}_{\rm SM}$ is the Standard Model Lagrangian, ${\cal L}_{\rm Clockwork}$ is the part of the Lagrangian involving only 
the new fermion singlets, and ${\cal L}_{\rm int}$ is the interaction term of the new fields with the Standard Model fields. 
Following \cite{Giudice:2016yja}, we assume that the Standard Model only couples to the last site of the fermionic clockwork, therefore, 
\begin{equation}
{\cal L}_{\rm int}
= - Y \widetilde H {\overline L}_L \psi_{R n} \;,
\label{eq:L_int}
\end{equation}
with $\widetilde H=i\tau_2 H^*$, $H$ the Standard Model Higgs doublet and ${L}_L$ the left handed lepton fields 
(we assume only one generation of fermions; the generalization to more than one generation will be discussed below).

In full generality, the clockwork Lagrangian can be cast as:
\begin{equation}
\mathcal{L}_{\rm Clockwork} =  \mathcal{L}_{\rm kin} -   \sum_{i=0}^{n-1}\left( m_i  {\overline \psi}_{L i} \psi_{R i} -m'_i\, {\overline \psi}_{L i} \psi_{R i\! + \! 1}
+ {\rm  h.c.} \right) 
-\sum_{i=0}^{n-1} \frac{1}{2}M_{Li} \overline{\psi^c_{Li}} \psi_{Li} 
-\sum_{i=0}^{n}\frac{1}{2}M_{Ri} \overline{\psi^c_{Ri}} \psi_{Ri}\;,
\label{lagfer}
\end{equation}
where $\mathcal{L}_{\rm kin}$ denotes the kinetic term for all fermions, and $m$, $m'$ and $M_{L,R}$ are mass parameters. Denoting $\Psi= (\psi_{L 0}, \psi_{L 1},... \psi_{L n-1}, \psi_{R 0}^{c},\psi_{R 1}^{c}, ...,\psi_{R n}^{c} )$, the clockwork Lagrangian can be written in the compact form:
\begin{equation}
\mathcal{L}_{\rm Clockwork} =  \mathcal{L}_{\rm Kin} - \frac{1}{2}(\overline{\Psi^c} {\cal{M}} \Psi + {\rm h.c.})
\label{ClockLag}
\end{equation}
with  ${\cal M}$ a $(2n+1)\times (2n+1)$ mass matrix. 
We note that ${\cal L}_{\rm kin}$ is invariant under the global group $U(n)_L\times U(n+1)_R$. 
The mass terms $m_i$ break the global group  $U(n)_L\times U(n+1)_R\rightarrow\prod_{i=0}^{n-1} U(1)_{i}$, where $U(1)_i$ acts 
as  $\psi_{L,i}\rightarrow e^{i\alpha_i}\psi_{L,i}$, $\psi_{Ri}\rightarrow e^{i\alpha_i}\psi_{Ri}$, and combined with the mass 
terms $m'_i$, break the global symmetry $U(n)_L\times U(n+1)_R\rightarrow U(1)_{\rm CW}$, where $U(1)_{\rm CW}$ acts 
as $\psi_{L,i}\rightarrow e^{i\alpha}\psi_{L,i}$, $\psi_{R,i}\rightarrow e^{i\alpha}\psi_{R,i}$ for all $i$. 
Finally, $M_{Li}$ and $M_{Ri}$ are Majorana masses for the left and right handed singlet fields. 
It is sufficient that $M_{Li}$ or $M_{Ri}$ is non-vanishing for one $i$ to break the symmetry group $U(n)_L\times U(n+1)_R\rightarrow$ nothing. 

We assume for simplicity universal Dirac masses, Majorana masses and nearest neighbor interactions, 
namely $m_{i} = m$, $m'_{i} = mq$ $M_{Ri} = M_{Li} = m \widetilde{q}$ for all $i$. Under this assumption, the mass matrix reads:
\begin{eqnarray}
{\cal{M}}= m
\begin{pmatrix}
\widetilde{q} & 0  
&\cdots & 0 &   1        & -q        &\cdots
& 0 \cr
0 & \widetilde{q}  
&\cdots & 0 &    0        & 1        &\cdots
& 0 \cr
\vdots   
&\vdots & \vdots & \vdots   & \vdots       & \vdots        &\vdots
& \vdots \cr
0 & 0 
&\cdots & \widetilde{q} & 0   & 0          &0
& -q \cr
1 & 0  
&\cdots & 0    & \widetilde{q}        & 0        &\cdots
& 0 \cr
-q & 1  
&\cdots & 0   &   0        & \widetilde{q}       &\cdots
& 0	\cr
\vdots   
&\vdots & \vdots & \vdots   & \vdots       & \vdots        &\vdots
& \vdots \cr
 0 & 0 
&\cdots & -q & 0 & 0       &0
&  \widetilde{q} \cr
\end{pmatrix}\;, 
\label{eq:general_mass_matrix}
\end{eqnarray}
which has eigenvalues $M_k$ given by:
	\begin{align}
	M_{0} &= m\widetilde{q}\;, \nonumber\\
	M_{k} &= m\widetilde{q} - m \sqrt{\lambda_k}\;, ~~~~k =1,\dots,n \;,  \nonumber\\
	M_{n+k} &= m\widetilde{q} + m \sqrt{\lambda_k}\;, ~~~~k =1,\dots,n\;,
	\label{GearMassMaj}
	\end{align}
with $\lambda_k$ defined as
\begin{equation}
\lambda_k \equiv q^2 + 1 -2q \cos \frac{k\pi}{n\! +\! 1} \;.
\label{lambdak}
\end{equation}

The mass eigenstates, which we denote as $\chi_k$, are related to  the interaction eigenstates $\Psi_j$ by the unitary transformation 
${\cal U}$, namely $\Psi_j= \sum_{j}{\cal U}_{jk} \chi_k$. The matrix ${\cal U}$ can be explicitly calculated, the result being:
	\begin{equation}
	{\cal U} = 
	\begin{pmatrix}
	\vec 0 &  \frac{1}{\sqrt{2}}U_L & - \frac{1}{\sqrt{2}}U_L \cr 
	\vec{u}_R & \frac{1}{\sqrt{2}}U_R & \frac{1}{\sqrt{2}}U_R\cr
	\end{pmatrix} \;.
	\end{equation}
	where $\vec 0$ and $\vec{u}_R$ are $n$-dimensional vectors, with entries:
	\begin{align}
	\vec 0_j &=0\;,~~~~ j=1, ..., n\;, \\
	(u_{R})_j  &=\frac{1}{q^j}\sqrt{\frac{q^2-1}{q^2-q^{-2n}}}  \;,~~~~  j=1, ..., n\;,
	\end{align}
	while $U_L$ and $U_R$ are, respectively, $n\times n$ and $(n+1)\times n$ matrices with elements
\begin{align}
(U_L)_{jk} &= \sqrt{\frac{2}{n+1}} \sin \frac{jk\pi}{n+1}\;, ~~~~~~~~ j,k =1,... , n\;, \nonumber\\
(U_R)_{jk} &=  \sqrt{\frac{2}{(n\! +\! 1)\lambda_k}} \left[ q \sin \frac{j k\pi}{n\! +\! 1}-  \sin \frac{(j +1) k\pi}{n\! +\! 1} \right] \, ,~~~ j =0, .. , n,~~k =1, ... , n\;,
\label{e:rotation}
\end{align}
We note that the mixing matrix ${\cal U}$ does not depend on the parameter $\widetilde q$, which is a consequence of our assumption of universality 
of the Majorana masses $M_{Ri}=M_{Li}=m\widetilde q$ for all $i$.

The interaction Lagrangian of the clockwork fields to the Standard Model fields, Eq.~(\ref{ClockLag}), can now be recast in terms of mass eigenstates:
\begin{align}
{\cal L}_{\rm int}
&= - Y \, {\overline L}_L \widetilde H {\cal U}_{nk}\chi_k \equiv -\sum_{k=0}^{2n} Y_k \, {\overline L}_L \widetilde H \chi_k \;,
\end{align}
where
\begin{align}
Y_0 &\equiv Y (u_R)_n =\frac{Y}{q^n}\sqrt{\frac{q^2-1}{q^2-q^{-2n}}} \;,  \\
Y_k=Y_{k+n} &\equiv \frac{1}{\sqrt{2}}Y (U_R)_{nk}=   Y
\sqrt{\frac{1}{(n\! +\! 1)\lambda_k}} \left[ q \sin \frac{n k\pi}{n\! +\! 1}\right] \;, ~~~~~ k=1, ..., n  \;.
\label{fkoneGen}
\end{align}
The components $(u_{R})_n$ and $(U_R)_{np}$, which describe the fraction of the $n_{th}$ ``gear'' in the zero mode, will play a major role 
in the phenomenology, as they parametrize the portal strength between the Standard Model sector and the clockwork sector. 

 After electroweak symmetry breaking new mass terms arise which mix the Standard Model neutrino with the clockwork fermions. 
 The mass matrix of the $2n+2$ electrically neutral fermion fields of the model reads:
\begin{eqnarray}
m_{\nu} =
\bordermatrix{
	&\nu_{L} & \chi_{0} 
	& \chi_{1} & \chi_{2} & \cdots 
	& \chi_{2n} \cr
	\nu_{L} & 0 & v Y_{0}  
	& v Y_{1} & v Y_{2} &   \cdots
	& v Y_{2n} \cr
	\chi_{0} & v Y_{0} & M_0
	& 0 &    0        &\cdots
	& 0 \cr
	\chi_{1} & v Y_{1} & 0 & M_1 
	& 0 \cr
	\chi_{2} & v Y_{2} & 0 & 0 & M_2  
	&\cdots
	& 0 \cr
	\vdots &\vdots   
	&\vdots & \vdots & \vdots   & \vdots       & \vdots \cr
	\chi_{2n} & v Y_{2n} & 0 & 0 & 0  &\cdots & M_{2n} 
	\cr
}\;,
\label{eq:general_case}
\end{eqnarray}	
where $v= 246/\sqrt{2}$ GeV is the Higgs vacuum expectation value.
Upon diagonalizing this mass matrix, one finds a mass for the active neutrino.  Furthermore, the off-diagonal entries in the mass matrix 
translate into charged current interactions between the charged lepton and the $k$-th mode, as well as neutral-current and Higgs interactions 
of the light neutrino, proportional to $\sim v Y_k/M_k$, and which can be sizable.

In order to accommodate the leptonic mixing observed in Nature it is necessary to introduce three generations of lepton doublets, 
as well as $N$ generation of clockwork fermions, each consisting of $n_\alpha$ left-handed and $n_\alpha+1$ right-handed gears, 
where $\alpha=1,\ldots, N$ (phenomenologically, $N\geq 2$, in order to account for the two observed oscillation frequencies).  
Furthermore, the Yukawa coupling in Eq.~(\ref{eq:L_int}) and all the mass parameters in Eq.~(\ref{lagfer}) must be promoted to 
matrices in flavor space. In this work we will assume for simplicity 
$m^{\alpha\beta}_{i} = m \delta^{\alpha\beta}$, ${m'}^{\alpha\beta}_{i} = mq_\alpha \delta^{\alpha\beta}$ $M^{\alpha\beta}_{Ri} = M^{\alpha\beta}_{Li} = m \widetilde{q}_\alpha \delta^{\alpha\beta}$ for all $i$. 
Namely, the mass parameter $m$ is universal for all gears and all generations, while the mass parameters $m'$, $M_{R}$ and $M_L$ are 
common for all gears within one generation, but in principle different among generations.

Denoting  $\Psi^\alpha= (\psi^\alpha_{L 0}, \psi^\alpha_{L 1}, ..., \psi^\alpha_{L n-1}, \psi_{R 0}^{\alpha\,c},\psi_{R 1}^{\alpha\,c}, ...,\psi_{R n}^{\alpha\,c} )$ as
the fermion field which has as component all the clockwork fields within the generation $\alpha$, the clockwork and interaction Lagrangian can be written as:
\begin{align}
\mathcal{L}_{\rm Clockwork} &=  \mathcal{L}_{\rm Kin} - \frac{1}{2}({\overline {\Psi^{\alpha\,c}}} {\cal{M}}^{\alpha\beta} \Psi^\beta + {\rm h.c.}) \;,\\
{\cal L}_{\rm int}
&= - Y^{a\alpha} \, \overline{L^a_L} \widetilde H \psi_{R,n}^\alpha \;,
\end{align}
where $a=1,2,3$ and $\alpha,\beta=1, ...,N$. As for the one generation case, we assumed that the Standard Model lepton doublets only couple 
to the $n$-th sites of the $N$ clockwork generations.

The Lagrangian expressed in the mass eigenstate basis, $\Psi^\alpha_{k}= {\cal U}^{\alpha\beta}_{kj}\chi^\beta_j$, read:
\begin{align}
\mathcal{L}_{\rm Clockwork} &=  \mathcal{L}_{\rm Kin} - \frac{1}{2}({\overline {\chi^{\alpha\,c}_k}} M^\alpha_k \chi^\alpha_k+ {\rm h.c.})\;, \\
{\cal L}_{\rm int}
&= - \sum_{k=0}^{2n} Y^{a\beta}_k \, {\overline L^a_L} \chi^\beta_k \;,
\end{align}
where $Y^{a\beta}_k \equiv Y^{a\alpha} {\cal U}_{nk}^{\alpha\beta}$ with ${\cal U}_{nk}^{\alpha\beta}$ the 
matrix that mixes fermions of different clockwork gears and different generations. Finally, after electroweak symmetry breaking, 
the mass matrix of the $N(2n+1)+3$ electrically neutral fermions of the model reads: 
	\begin{eqnarray}
	m_{\nu} =
	\bordermatrix{
		&\nu^a_{L} & \chi^\beta_{0} 
		& \chi^\beta_{1} & \chi^\beta_{2} & \cdots 
		& \chi^\beta_{2n} \cr
		\nu^a_{L} & 0 & v Y^{a \beta}_{0}  
		& v Y^{a \beta}_{1} & v Y^{a \beta}_{2} &   \cdots
		& v Y^{a \beta}_{2n} \cr
		\chi^\beta_{0} & v Y^{\beta a}_{0} & M_0^{\beta}
		& 0 &    0        &\cdots
		& 0 \cr
		\chi^\beta_{1} & v Y^{\beta a}_{1} & 0 & M_1^{\beta} 
		& 0 \cr
		\chi^\beta_{2} & v Y^{\beta a}_{2} & 0 & 0 & M_2^{\beta}
		&\cdots
		& 0 \cr
		\vdots &\vdots   
		&\vdots & \vdots & \vdots   & \vdots       & \vdots \cr
		\chi^\beta_{2n} & v Y^{\beta a}_{2n} & 0 & 0 & 0  &\cdots & M_{2n}^{\beta}
		\cr
	}\ \;.
	\end{eqnarray}	
This matrix has in general a non-trivial flavor structure and leads not only to mixing among the three active neutrinos, 
but also to potentially large lepton flavour violating charged current, neutral current and Higgs interactions, thus providing a 
possible test of this framework, as will be discussed in Section \ref{LeptFlav}.

We consider in what follows two cases: $M_{Li}, M_{Ri}=0$, for all $i$, such that the Clockwork Lagrangian has a 
residual $U(1)_{\rm CW}$ global symmetry, and  $M_{Li}, M_{Ri}\neq 0$ for some $i$, such that the Clockwork Lagrangian has no global symmetry.

\subsection{$M_{Li}, M_{Ri}=0$, for all $i$}\label{Dirac}

We consider first the case where all Majorana masses are equal to zero. In this case, the global symmetry of the 
Lagrangian is broken as  $U(n)_L\times U(n+1)_R\rightarrow U(1)_{\rm CW}$, which will be identified with total lepton number. 
The eigenstates and eigenvalues of the mass matrix can be determined using the results of Section \ref{neutrino}, by setting $\widetilde q=0$. 

\begin{figure}[tbp]
	\centering
	\includegraphics[height=5.8cm, width=8.0cm]{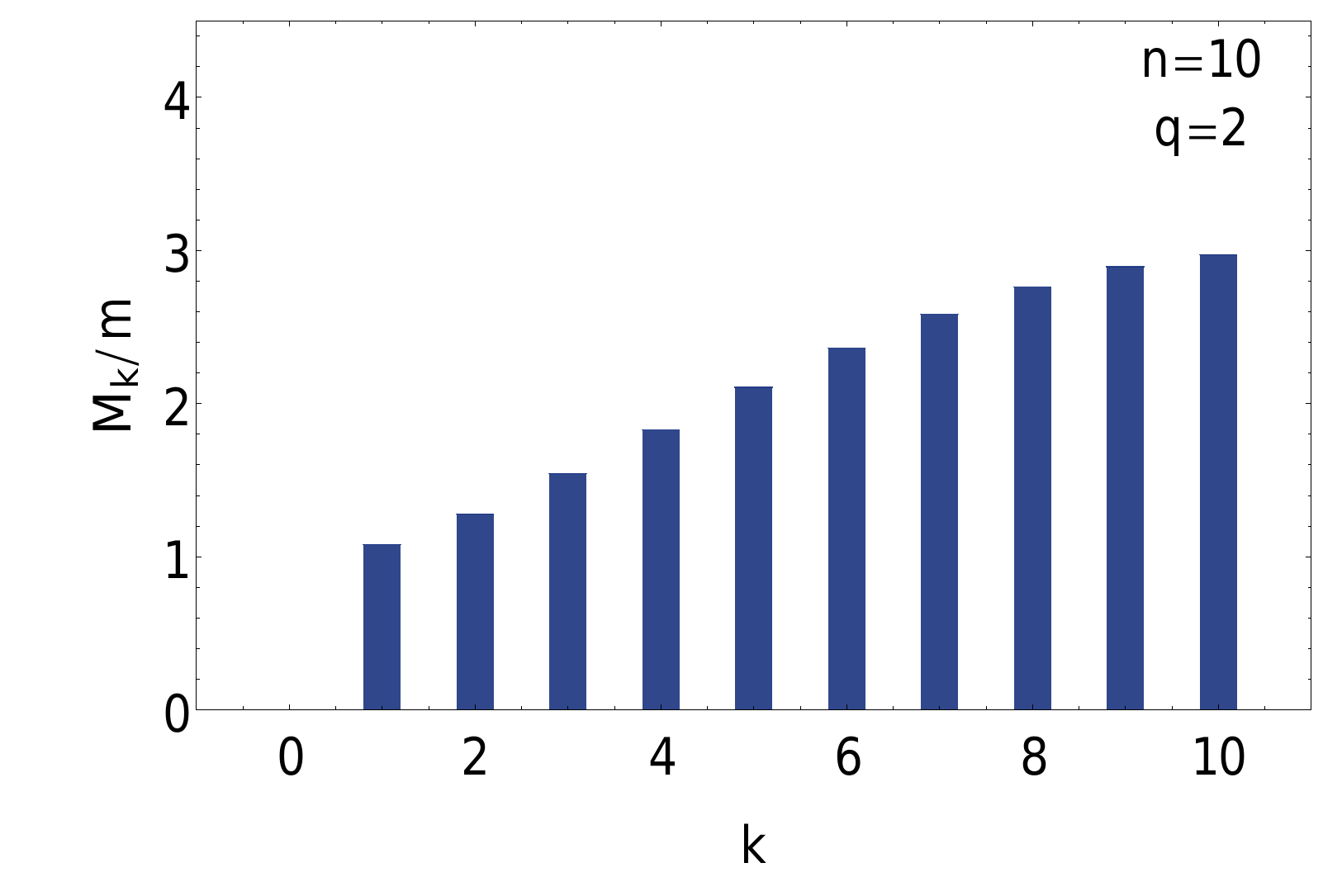}~~
	\includegraphics[height=5.8cm, width=8.cm]{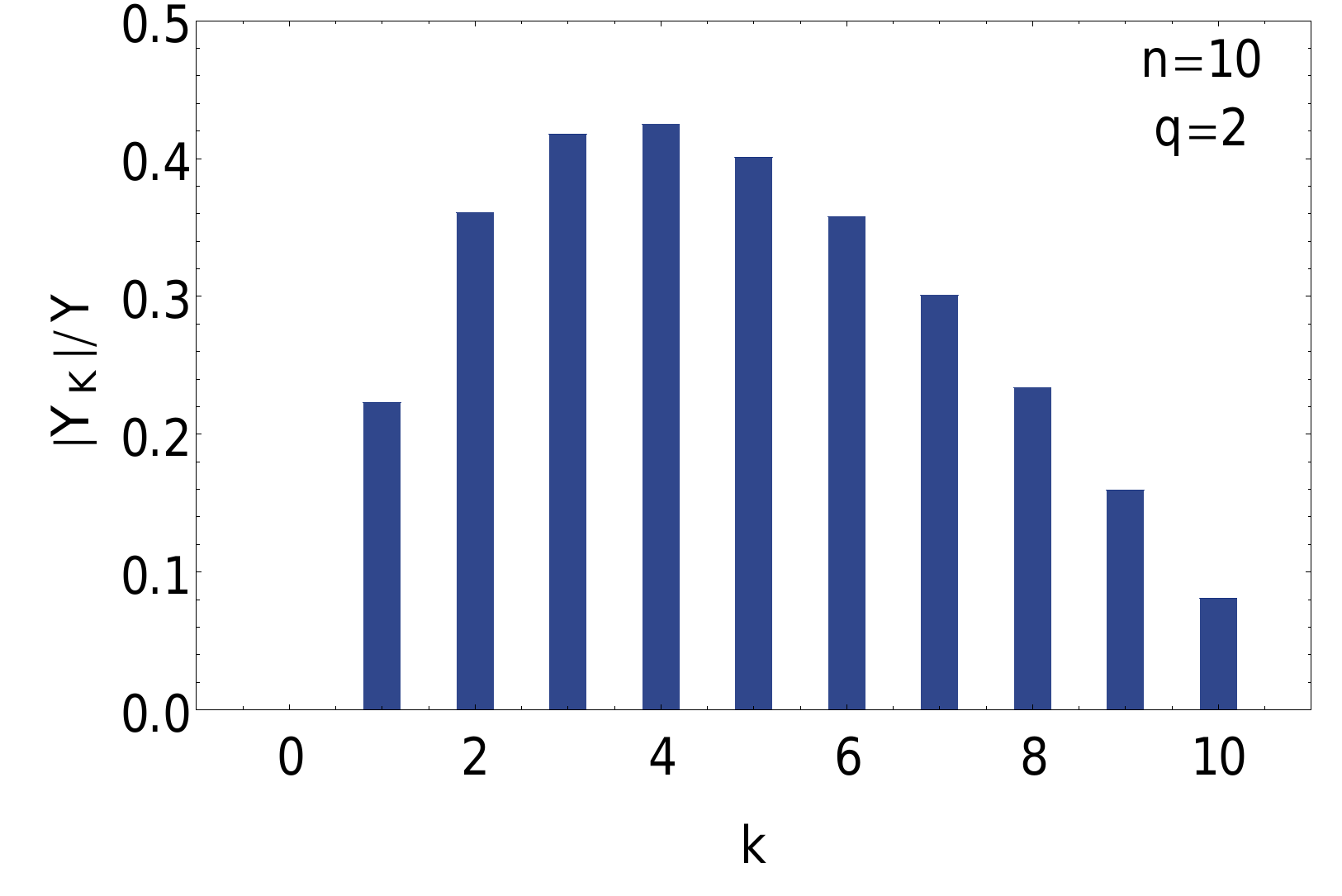}
	\caption{\sf Dirac masses (left panel) and Yukawa couplings (right panel) of the singlet fermions of the clockwork sector, normalized respectively to $m$ and $Y$, for the specific case $n=10$ and $q=2$.}
	\label{fig:Dirac}
\end{figure}

It is useful to recast the clockwork Lagrangian as
\begin{align}
{\mathcal L}_{\rm clockwork} &={\mathcal L}_{\rm kin} -
\overline{N_{L}} m^D_{\nu} N_{R}+ {\rm h.c.} 
\end{align}
where we have defined new fields
$N_{L} = (\nu_{L}, N_{L1}, ...,N_{L n})$ and 
$N_{R} = (N_{R0}, N_{R1}, ...,N_{R n})$, with
\begin{align}
N_{Rk}&=\frac{1}{\sqrt{2}}(\chi_{k}+\chi_{k+n})\;,~~~~~k=0, ... n\;,\\	
N_{Lk}&=\frac{1}{\sqrt{2}}(-\chi_{k}+\chi_{k+n})\;,~~~~~k=1, ... n\;.
\end{align}
In this basis, the mass matrix has the form:
\begin{eqnarray}
m^D_{\nu} =
\bordermatrix{
	&N_{R0} & N_{R1} & N_{R2}	& \cdots &  N_{Rn} \cr
	\nu_{L} &  0 & 0& 0 &   \cdots	&0 \cr
	N_{L1} & 0 & M_1 &0 &\cdots& 0 \cr
	N_{L2} & 0 & 0 & M_2 &\cdots& 0 \cr
	\vdots &\vdots 	&\vdots &\vdots  &\ddots & \vdots \cr
	N_{Ln} & 0 & 0  &0   &\cdots & M_{n}	\cr
}\ \;.
\end{eqnarray}	
where $M_{k} = m \sqrt{\lambda_k}$, with $\lambda_k$ defined in Eq.~(\ref{lambdak}). 
Namely, the fields $\nu_L$ and $N_{R0}$ form a massless Dirac pair, while the fields $N_{Rk}$ and $N_{Lk}$ form, for $k=1, ..., n$, 
Dirac pairs with mass $M_k$. The overall scale of the massive pairs is determined by the parameter $m$, and the mass difference 
between pairs depends on $q$ and $n$. Assuming $q>1$, one obtains that the masses of the modes with $k>0$ increase monotonically 
with $n$, from $M_{1} \approx m(q-1)$ to $M_{n} \approx m(q+1)$. In Fig.~\ref{fig:Dirac}, left panel, we show for illustration 
the mass spectrum of the particles of the clockwork sector, labeled by $k$, taking for concreteness $n=10$ and $q=2$. 
The mass spectrum has been normalized to $m$.

The mass spectrum is modified after electroweak symmetry breaking by the interactions with the Higgs field. 
Expressed in terms of $N_{Rk}$, the interaction Lagrangian reads:
\begin{align}
{\mathcal L}_{\rm int} &=\sum_{k=0}^n Y_k{\overline L}_L \widetilde{H} {N_{R k}}+{\rm h.c.}
\label{diracLag}
\end{align}
with
\begin{align}
Y_0 &\equiv Y (u_R)_n =\frac{Y}{q^n}\sqrt{\frac{q^2-1}{q^2-q^{-2n}}}  \;,~~~~  \\
Y_k &\equiv Y (U_R)_{nk}= Y
\sqrt{\frac{2}{(n\! +\! 1)\lambda_k}} \left[ q \sin \frac{n k\pi}{n\! +\! 1}\right] \;,~~~~~ k=1, ..., n\;.
\label{fkoneGen}
\end{align}
The Yukawa coupling of the massless mode $Y_0$ is suppressed by $q^n$, provided $q>1$, 
whereas the couplings of the $k$th-mode are of the same order as $Y$. This is illustrated in Fig.~\ref{fig:Dirac}, 
right panel, which shows the Yukawa couplings of the clockwork fermions to the Standard Model lepton doublets, 
normalized to $Y$, for the same values of $n$ and $q$ as in the left panel (in this case, $|Y_0|/Y\approx 8\times 10^{-4}$ 
and is not visible from the figure.)

The mass matrix of the electrically neutral fermion fields now reads:
\begin{eqnarray}
m^D_{\nu} =
\bordermatrix{
	&N_{R0} & N_{R1} & N_{R2}	& \cdots &  N_{Rn} \cr
	\nu_{L} &  v Y_{0} & v Y_{1}& v Y_{2} &   \cdots	& v Y_{n} \cr
	N_{L 1} & 0 & M_1 &0 &\cdots& 0 \cr
	N_{L2} & 0 & 0 & M_2 &\cdots& 0 \cr
	\vdots &\vdots 	&\vdots &\vdots  &\ddots & \vdots \cr
	N_{Ln} & 0 & 0  &0   &\cdots & M_{n}	\cr
}\ \;.
\end{eqnarray}	
 Concretely, a mass term for the active neutrinos is generated. Assuming that $M_k\gg Y_0 v$, which as we will see 
 below is justified from the current limits on rare leptonic decays, one can approximate the active neutrino mass by 
\begin{align}
m_{\nu}\approx v Y_0
\end{align}
and can be made small by choosing appropriate values of $Y$, $q$ and $n$. For instance, assuming $Y={\cal O}(1)$, $q=2$, 
one obtains $m_\nu ={\cal O}(0.1)\,{\rm eV}$ for $n\approx 40$.

The generalization of the above setup to three leptonic generations and $N$ clockwork generations is straightforward. The clockwork Lagrangian is:
\begin{align}
{\mathcal L}_{\rm clockwork} &={\mathcal L}_{\rm kin} -
\overline{N^\alpha_{L}} m^\alpha_{\nu} N^\alpha_{R}+ {\rm h.c.} 
\end{align}
with
$N^\alpha_{L} = (\nu^\alpha_{L}, N^\alpha_{L1}, ...,N^\alpha_{Ln})$ and 
$N^\alpha_{R} = (N^\alpha_{R0}, N^\alpha_{R1}, ...,N^\alpha_{Rn})$,
where 
\begin{align}
N^\alpha_{Rk}&=\frac{1}{\sqrt{2}}(\chi^\alpha_{k}+\chi^\alpha_{k+n})\;,~~~~~k=0, ..., n\,~~\alpha=1...,N\;,\\	
N^\alpha_{Lk}&=\frac{1}{\sqrt{2}}(-\chi^\alpha_{k}+\chi^\alpha_{k+n})\;,~~~~~k=1, ..., n,~~\alpha=1...,N\;,
\end{align}
and the interaction Lagrangian,
\begin{align}
{\mathcal L}_{\rm int} =- \sum_{k=0}^n Y_k^{a\beta}\overline{L_L^a} \widetilde{H}_0 N_{Rk}^\beta\;,
\end{align}
with $Y_k^{a\beta}= Y^{a\alpha}{\cal U}^{\alpha\beta}_{nk}$.

After electroweak symmetry breaking the neutrino mass matrix reads:
\begin{eqnarray}
m^D_{\nu} =
\bordermatrix{
	&N^\beta_{R0} & N^\beta_{R1} & N^\beta_{R2}	& \cdots &  N^\beta_{Rn} \cr
	\nu^a_{L} &  v Y^{a \beta}_{0} & v Y^{a \beta}_{1}& v Y^{a \beta}_{2} &   \cdots	& v Y^{a \beta}_{n} \cr
	N^\beta_{L1} & 0 & M_1^{\beta} &0 &\cdots& 0 \cr
	N^\beta_{L2} & 0 & 0 & M_2^{\beta} &\cdots& 0 \cr
	\vdots &\vdots 	&\vdots &\vdots  &\ddots & \vdots \cr
	N^\beta_{Ln} & 0 & 0  &0   &\cdots & M_{n}^{\beta}	\cr
}\ \;.
\end{eqnarray}	
where $M_{k}^\beta$ is the mass of $k$-th clockwork gear for the Dirac pair $N_L^\beta$,$N_R^\beta$.

\begin{figure}[tbp]
	\centering
	\includegraphics[height=5.8cm, width=7.9cm]{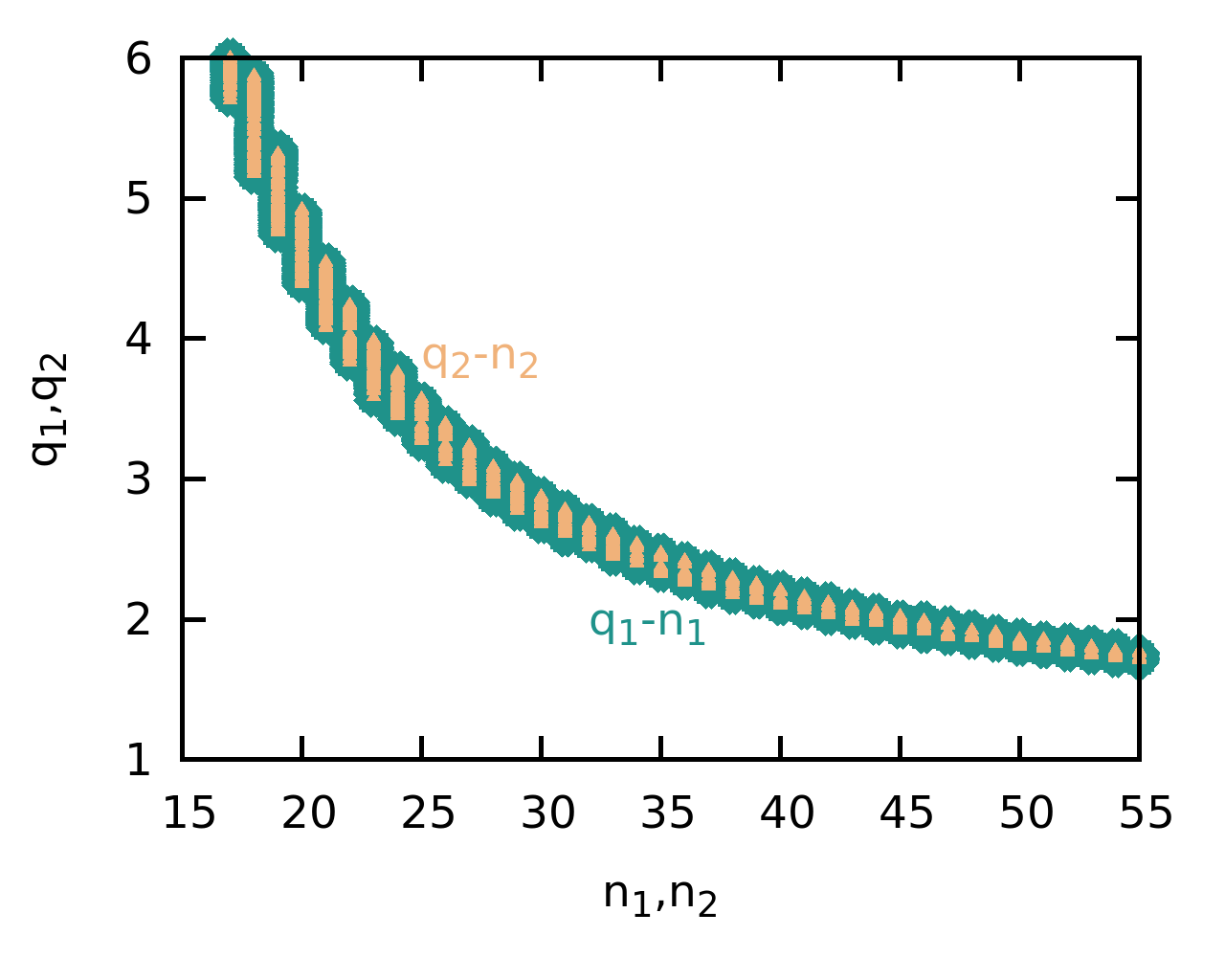}
	\includegraphics[height=5.8cm, width=7.9cm]{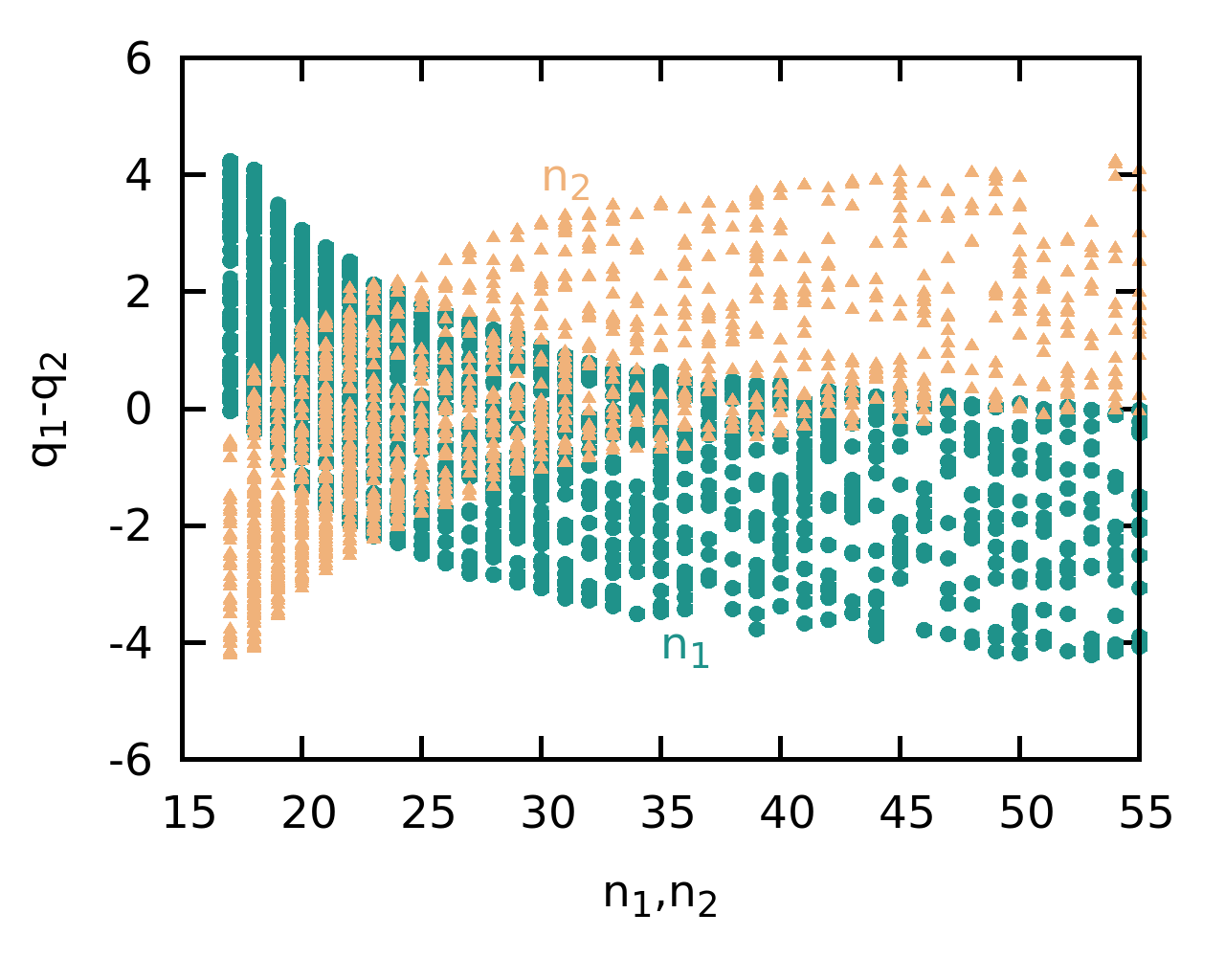}
	\caption{\sf Values of $q_1$ and $q_2$ (left panel) and difference between them (right panel), as a function of $n_1$ and $n_2$, compatible with the measured values of the neutrino mass splittings and mixing angles within $1\sigma$, for a scenario with two clockwork generations.}
	\label{fig:two_gen_CW}
\end{figure}

We analyze in detail the case where the clockwork consists of two generations with $n_1$ and $n_2$ gears, respectively. 
We scan $Y^{a\alpha}$ within the ranges $\frac{1}{4}<| Y^{a\alpha}|<4$, $q_\alpha$ between 1.5 and 6 and $n_\alpha$ between 15 and 55, 
and we select the points that reproduce the observed values of the solar and atmospheric mass splitting and mixing angles within $1\sigma$, 
as determined in Ref.~\cite{Capozzi:2017ipn}. In Fig.~\ref{fig:two_gen_CW} (left panel) we show as green circles (yellow triangles) 
the values of $n_1$ ($n_2$) as a function of $q_1$ ($q_2$)  that satisfy the experimental constraints. As apparent from the plot, 
larger $q_\alpha$ require a smaller number of gears to reproduce the small neutrino Yukawa coupling. 
Furthermore, the allowed values for $n_1$ and $n_2$ have a big overlap, which is a consequence of our assumption of comparable 
elements in the coupling $Y^{a\alpha}$ and the necessity of producing a mild hierarchy between the solar and the atmospheric 
neutrino mass scales. In particular, we find that the scenario with $q_1=q_2$ and $n_1=n_2$, namely the scenario where the 
clockwork parameters are universal also among generations, is allowed by observations. 
This is illustrated in Fig.~\ref{fig:two_gen_CW} (right panel), which shows the allowed values of $q_{1}-q_{2}$ as a 
function of $n_{1}$(green circle) and $n_{2}$(yellow triangle); the scenario with $n_{1}=n_{2}$ and $q_{1}=q_{2}$ corresponds 
to the region where the green circles and the yellow triangles overlap. 

\subsection{$M_{Li}, M_{Ri}\neq 0$ for some $i$}\label{Majorana}
In this case the mass matrix of the model is given by Eq.~(\ref{eq:general_case}) and the Yukawa couplings by Eq.~(\ref{fkoneGen}). 
Identifying $\widetilde q$ as the order parameter of the $U(1)_{\rm CW}$ symmetry breaking, one can consider two 
limits of interest: $\widetilde q\ll q,1$ and  $\widetilde q\gg q,1$. 

\begin{figure}[tbp]
	\centering
		\includegraphics[height=5.8cm,width=7.9cm]{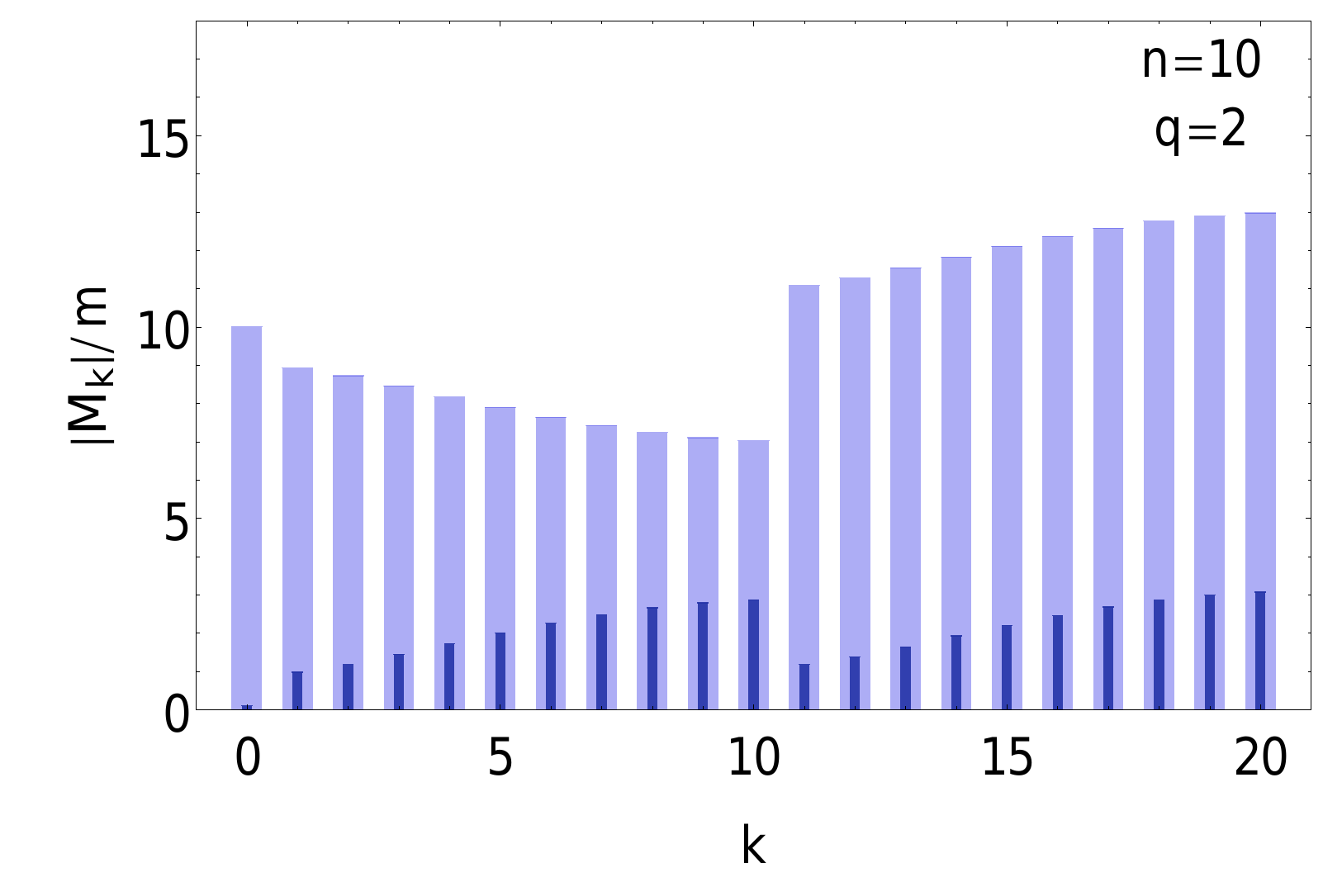}	\includegraphics[height=5.8
		cm, width=7.9cm]{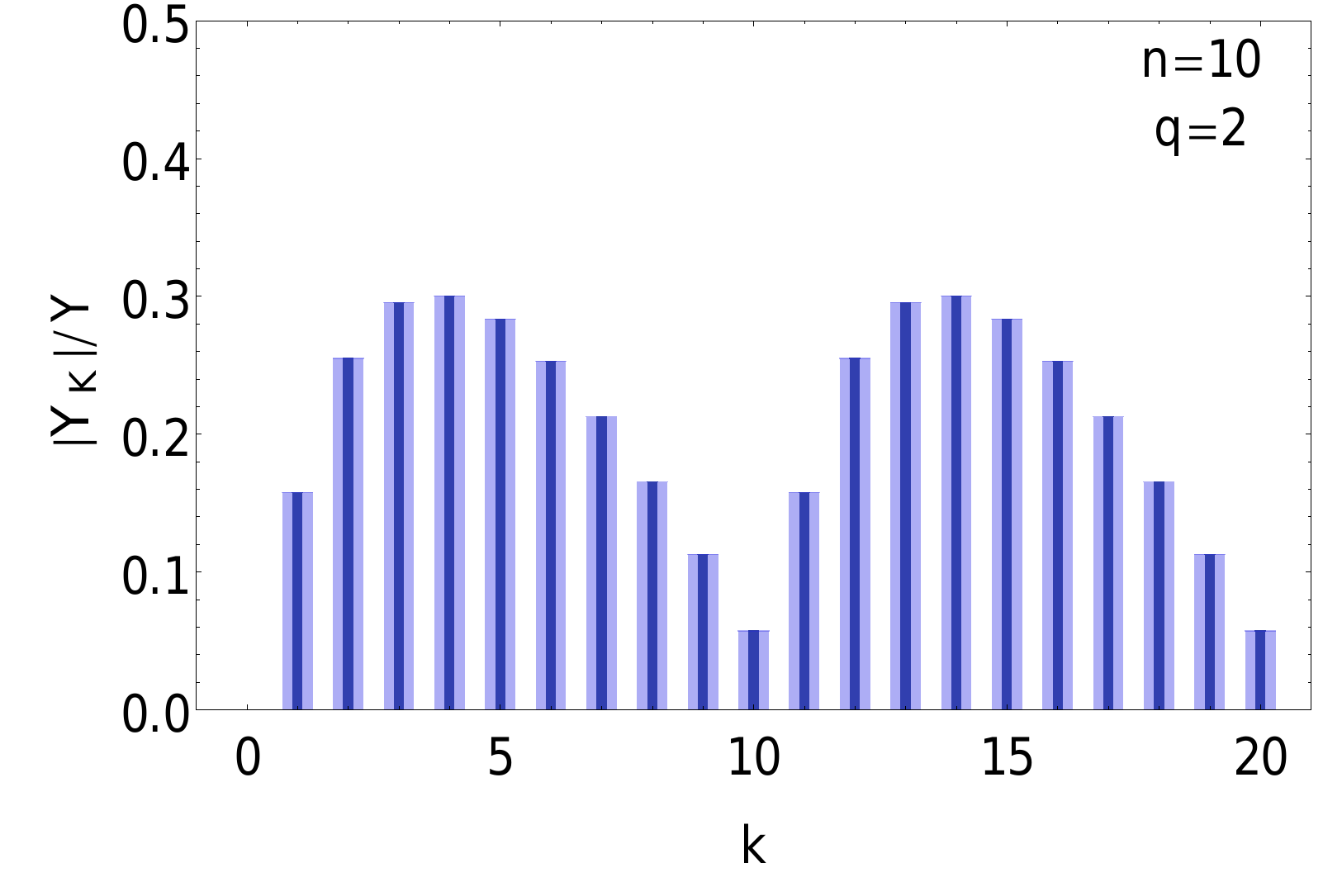}
		\caption{\sf Majorana masses (left panel) and Yukawa couplings (right panel) of the singlet fermions of the clockwork sector, normalized respectively to $m$ and $Y$, for the specific case $n=10$, $q=2$ and $\widetilde q=0.1$ (dark blue) or $\widetilde q=10$ (light blue).}
		\label{fig:Majorana}
\end{figure}

Fig. \ref{fig:Majorana} shows the masses of the singlet fermions (left panel) and their corresponding Yukawa couplings 
(right panel) for the specific case $n=10$, $q=2$, and $\widetilde q=0.1$ (dark blue) or $\widetilde q=10$ (light blue); the 
former case corresponds to a mild breaking of the $U(1)_{\rm CW}$ symmetry and the latter to a strong breaking. For $\widetilde q=0.1$ 
one notices that the mode $k$ and the mode $n+k$ have very similar masses and suggest a pseudo-Dirac structure, 
which results from the mild $U(1)_{\rm CW}$ breaking; in the limit $\widetilde q\rightarrow 0$, they would form an exact 
Dirac pair and have identical masses. For $\widetilde q=10$, however, the masses of all the modes are markedly different.

 On the other hand, the Yukawa couplings of the singlet fermions to the left-handed leptons, shown in the right panel, 
 do not depend on the value of $\widetilde q$, as demonstrated in subsection \ref{Dirac}. The phenomenology of the 
 scenario $\widetilde q\ll q,1$ is then very similar to the one already discussed in subsection  \ref{Dirac}, while 
 the phenomenology of the scenario $\widetilde q\gg q,1$ can be rather distinct from the one in the (pseudo-)Dirac case. 
 Indeed, in this scenario one obtains a mass for the active neutrino through the seesaw mechanism given by:
\begin{equation}
m_{\nu} \approx \sum_k \dfrac{Y_k^2v^2}{M_{k}} \;.
\label{seesaw} 
\end{equation}
Then, since the couplings for the higher modes are expected to be ${\cal O}(Y)$, the resulting neutrino mass can be orders of magnitude larger than the value inferred from oscillation experiments, unless $Y\ll 1$ and/or the gear masses are very large, in the same spirit as in the standard seesaw mechanism. A similar conclusion was also reached in  \cite{Park:2017yrn}.

\section{Lepton Flavor Violation}\label{LeptFlav}
The clockwork mechanism suppresses the Yukawa couplings for the zero mode, hence explaining the smallness of neutrino masses. However the Yukawa couplings for the higher modes are in general unsuppressed and can lead to observable effects at low energies. In particular, the lepton flavor violation generically present in the Yukawa couplings of the higher modes contributes, through quantum effects induced by clockwork fermions, to generate rare leptonic decays (such as $l_{i} \to l_{j} \gamma$) or $\mu$-e conversion in nuclei, with rates that could be at the reach of current or future experiments if the gear masses are sufficiently low. 

We calculate the rate for $l_{i} \to l_{j} \gamma$ following \cite{Petcov:1976ff,Bilenky:1977du, Cheng:1980tp}. For $N$ clockwork generations, we obtain:
\begin{equation}
B\left( \mu \rightarrow e\gamma \right)\simeq\frac{3\alpha_{\rm em} v^4}{8\pi }\left|
\sum_{\alpha=1}^N\sum_{k=1}^{n_{\alpha}}
\frac{Y_k^{e\alpha}Y_k^{\mu \alpha}}{{M^\alpha_{k}}^2} F(x_k^\alpha)\right| ^{2},  \notag 
\end{equation}
where $\alpha_{\rm em}$ is the fine structure constant, $n_\alpha$ is the number of gears in the $\alpha$-th generation, 
$M_k^{\alpha}$ is the mass of the $k$-th mode in the  $\alpha$-th generation ($k=1, ..., n_\alpha$), 
and $x_k^\alpha \equiv {M_k^\alpha}^2/M^2_W$. The loop function $F(x)$ is defined as 
\begin{eqnarray}
F(x)&\equiv&\frac{1}{6(1- x)^{4}}
(10-43 x +78 x^2 -49 x^3
+4 x^4
-18 x^3 \log x)\ , 
\end{eqnarray}
and has limits $F\left( 0\right) =5/3$ and $F\left( \infty \right) =2/3$. 

\begin{figure}[t]
	\centering
	\includegraphics[height=5.9cm, width=8.cm]{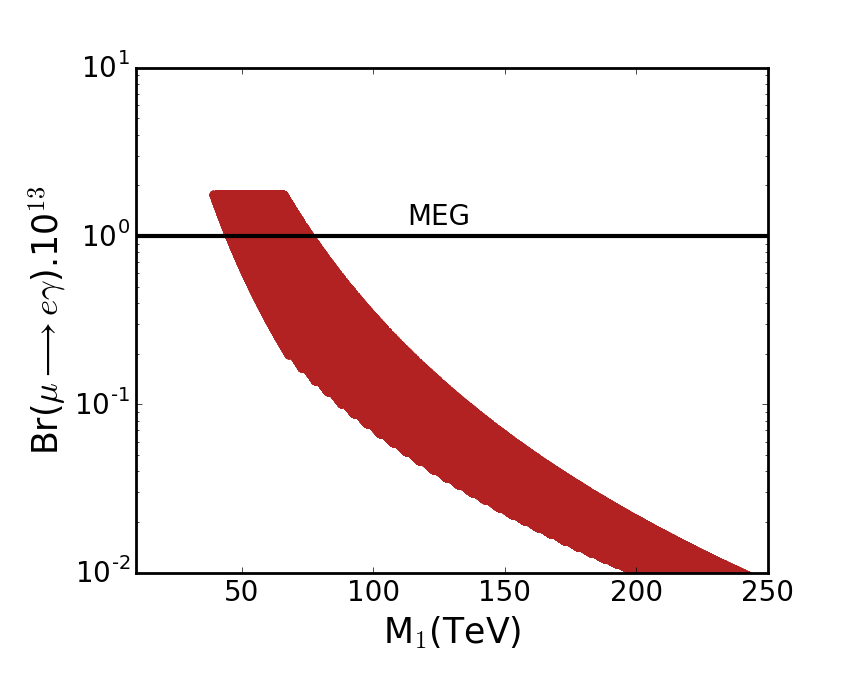}
		\caption{\sf Predicted value of $Br(\mu\rightarrow e \gamma)$ for points of the parameter space reproducing the observed neutrino oscillation parameters, as a function of the mass of the first clockwork gear. The black solid line shows the current upper limit from the MEG experiment.}\label{fig:mueg}
\end{figure}

The current upper bound ${\rm Br}(\mu\rightarrow e \gamma)\leq 4.2\times10^{-13}$ from the MEG experiment\cite{TheMEG:2016wtm}
poses stringent constraints on the mass scale of the clockwork. In Fig.\ref{fig:mueg} we show the  branching ratio expected 
for points reproducing the measured neutrino parameters, assuming two clockwork generations, as obtained in the scan  
presented in section \ref{Dirac}, as a function of the mass of the first clockwork gear. It follows from the figure that the 
clockwork gears must be larger than $\sim$ 40 TeV in order to evade the experimental constraints, unless very fine cancellations 
among all contributions to this process exist. For a larger number of clockwork generations we expect even stronger lower 
limits on the lightest gear mass, due to the larger number of particles in the loop.

\section{Summary}

The origin of small neutrino masses remains a mystery to this day. The recently proposed clockwork mechanism provides new insights into this puzzle, as it naturally generates small parameters in the effective Lagrangian. In the present work, we have scrutinized the mechanism of neutrino mass generation within the clockwork framework. We have generalized the clockwork formalism to include, in addition to Dirac masses and nearest neighbor interactions, also Majorana mass terms in the clockwork sector; and we have derived analytical expressions for the masses and couplings of the new singlet fermions for the specific case where the Dirac masses, Majorana masses and nearest neighbor interactions are universal among all clockwork ``gears". 

We have investigated in detail the impact of the Majorana masses in the clockwork sector in the generation of small neutrino masses. When the Majorana masses vanish, the zero mode of the clockwork sector is strictly massless and forms a Dirac pair with the active neutrino. In this framework, small Dirac neutrino masses can be generated for a sufficiently large number of gears, depending on the hierarchy between the mass scales in the clockwork sector. On the other hand, when the Majorana masses are non-vanishing, the zero mode is no longer massless. However, the corresponding Yukawa coupling still has the clockwork structure. 
In this case, small neutrino masses are the result of the interplay between the standard seesaw mechanism and the ``clockworked" Yukawa couplings, and typically require very large Majorana masses in order to reproduce the small neutrino mass scale inferred from oscillation experiments. 

The Standard Model leptons couple to the fermions of the clockwork sector with a site dependent strength, giving rise to (possibly lepton flavour violating) charged current, neutral current and Higgs boson interactions. We have investigated the constraints on this framework from the non-observation of the rare leptonic decay $\mu\rightarrow e\gamma$. Our results indicate that the lightest particle of the clockwork sector must have a mass $\gtrsim 40$ TeV, if the Yukawa couplings of the fundamental theory are ${\cal O}(1)$.

\section*{Acknowledgments}
 AI and SKV acknowledge partial financial support from the DFG cluster of excellence EXC 153 ``Origin and Structure of the Universe'' and AI from the Collaborative Research Center SFB1258. SKV thanks the Physics Department of the Technical University of Munich for hospitality. SKV thanks the hospitality of IPHT, CEA, Saclay during the final stages of this work.

\bibliographystyle{ieeetr}
\bibliography{Neutrino.bib}
\end{document}